\documentclass{PoS}

\title{Type IIn supernovae as sources of high energy neutrinos}

\ShortTitle{IIn supernovae as sources of high energy neutrinos}

\author{\speaker{V.N.Zirakashvili}\\
        Pushkov Institute of Terrestrial Magnetism, Ionosphere and Radiowave
Propagation, 142190, Troitsk, Moscow, Russia\\
        E-mail: \email{zirak@izmiran.ru}}

\author{V.S.Ptuskin\\
        Pushkov Institute of Terrestrial Magnetism, Ionosphere and Radiowave
Propagation, 142190, Troitsk, Moscow, Russia\\
        E-mail: \email{vptuskin@izmiran.ru}}

\abstract{
It is shown that high-energy astrophysical neutrinos observed in the IceCube experiment
can be produced
 by protons accelerated in extragalactic Type IIn supernova remnants by shocks
propagating in the dense circumstellar medium.
 The nonlinear diffusive shock acceleration model is used for description of
particle acceleration.
}

\FullConference{The 34th International Cosmic Ray Conference,\\
        30 July- 6 August, 2015\\
        The Hague, The Netherlands}

\begin{document}

\section{Introduction}

The detection of very high energy astrophysical neutrinos in the IceCube
experiment \cite{aartsen13, aartsen14} opens up a new possibility for
investigation of particle acceleration processes in the Universe.
The neutrino production in cosmos is possible via the $pp$ and
$p\gamma$ interactions and the decay chains
$\pi^{+}\rightarrow \mu^{+} \nu_{\mu}$, $\mu^{+}\rightarrow e^{+}\nu_{e}\bar{\nu_{\mu}}$.
The observed astrophysical flux emerges from under more steep air show spectrum at about
$50$ TeV and has a cutoff at $2$ PeV. The neutrino typically carries small part of the
primary proton energy, $E_{\nu}\approx 0.05E_{p}$, so the protons with energies up
to $E_{max}\sim 10^{17}$ eV are required to explain the observations (this energy
is $\sim 10^{17}$ eV/nucleon in the case of neutrino production by nuclei).
Assuming an $E^{-2}$ power-law spectrum, the measured differential flux of
astrophysical neutrinos is $E^{2}F(E)=2.9\times 10^{-8} \textrm{GeV}^{-2}\textrm{s}^{-1}\textrm{sr}^{-1}$
for the sum of the three evenly distributed neutrino flavors. The sources of observed neutrinos
are not yet identified. The detected $37$ events are scattered over the sky and do not show
evident correlation with any astronomical objects \cite{aartsen14b}. It seems that Galactic
sources might account only for a minority of events. The detected astrophysical neutrinos
could be produced in extragalactic sources of ultra high energy protons and nuclei. The
discussion about potential sources of very high energy neutrinos in the light of the
last experimental results can be found in \cite{Anchor14,GaisserHalz14,murase14a}
where other useful references are given.

The rare extragalactic Type IIn supernova remnants are considered in the present paper as
 sources of diffuse high energy neutrinos. It is well established that supernova remnants
are efficient accelerators of protons, nuclei and electrons. They are the principle sources
of Galactic cosmic rays. The diffusive shock acceleration mechanism suggested
in \cite{krymsky77,bell78,axford77,blandford78} can provide the acceleration
of protons and nuclei in the most frequent Type IIP, Ia, Ib/c supernova events
up to about $10^{15}Z$ eV that allows to explain the spectrum and composition
of Galactic cosmic rays with a proton-helium knee at $3\times 10^{15}$ eV and
the maximum energy $\sim 10^{17}$ eV where iron nuclei dominate,
see \cite{ptuskin10}. Two orders of magnitude larger $E_{max}\sim 10^{17}$ eV/nucleon
is needed to explain the IceCube neutrino observations. It can be achieved with the Type
IIn supernovae that stand out because of extremely dense wind of their progenitor stars with a mass loss rate
$10^{-3} - 10^{-1}\ M_{\odot}$ yr$^{-1}$ \cite{moriya14}. As it will be shown below, the
large kinetic energy of explosion and very high gas density in the acceleration region
lead to the needed energy of accelerated particles and efficiency of neutrino production in $pp$ interactions.

Diffusive shock acceleration  by supernova shocks propagating in dense stellar winds was already
considered in \cite{kats11,murase14}. Simple analytical estimates
showed that radiowaves, gamma-rays and neutrinos  might be observable from the nearest
Type IIn supernova remnants if the efficient diffusive shock acceleration takes place
in these objects. In the present paper we develop this idea further and investigate
whether the flux of neutrinos produced in extragalactic Type IIn supernova remnants
can explain the IceCube data. For this purpose we perform numerical modeling of
particle acceleration in
a supernova remnant produced by Type IIn supernova explosion and calculate neutrino
production. Our model of nonlinear diffusive shock
acceleration describes the remnant evolution and the production of energetic particles.
The detailed description of the model was presented in \cite{zirakashvili12} and the
simplified version of the model was used in \cite{ptuskin10} for the explanation of
energy spectrum and composition of Galactic cosmic rays. Similar numerical models of
diffusive shock acceleration in supernova remnants were developed and employed in
\cite{berezhko94,kang06,berezhko07}.

The paper is organized as follows. In the next Section 2  we
describe modeling of particle acceleration and calculate the
 spectrum of neutrinos produced in Type IIn supernova remnants.
These results are used in Section 3 for calculation of diffuse neutrino background in the expanding Universe.
The discussion of results and conclusions are given in Sections 4 and 5.

\section{Modeling of diffusive shock acceleration in the remnant of Type IIn supernova}

Details of our model of nonlinear diffusive shock acceleration can be found in \cite{zirakashvili12}.
The model contains coupled spherically symmetric
hydrodynamic equations  and
the transport equations for energetic protons, ions and electrons. The forward and reverse shocks are
included in the consideration.

The blast wave produced by Type IIn supernova explosion propagates through the wind of the
presupernova star. We assume that the initial stellar wind density profile is described by the following expression:

\begin{equation}
\rho =\frac {\dot{M}}{4\pi u_w\ r^2}.
\end{equation}
Here $\dot{M}$ is the mass-loss rate of the supernova progenitor star, $u_w$ is the wind velocity and $r$ is the
 distance from the center of explosion.

We use the following parameters of the supernova explosion. The explosion energy $E_{SN}=10^{52}$ erg,
 ejecta mass $M_{ej}=10\ M_{\odot}$, $\dot{M}=10^{-2}\ M_{\odot}$ yr$^{-1}$, the parameter of ejecta
velocity distribution $k=9$ (this parameter describes the power-law density profile $\rho_{ej}\propto r^{-k}$
of the outer part of the ejecta that freely expands after supernova explosion),
the stellar wind velocity $u_w=100$ km s$^{-1}$ that are the characteristic values for
Type IIn supernova \cite{moriya14}.

The injection efficiency of thermal protons $\eta $  is taken to be
independent of time $\eta =0.01$. Protons of mass $m$ are
injected at the forward and reverse shocks.
The high injection efficiency
results in the significant shock modification already at early stage of the supernova remnant expansion.

The secondary electrons and positrons are effectively produced in the
dense medium of Type IIn supernova remnant via
 {\it pp} interactions. That is why we do not take into account injection of thermal
electrons at the shocks in the present calculations.

Cosmic ray diffusion is determined by particle scattering on magnetic inhomogeneities. The cosmic ray 
streaming instability increases the
level of magnetohydrodynamic  turbulence in the shock vicinity \cite{bell78} and even 
significantly amplify the 
absolute value of magnetic field in young supernova remnants \cite{bell04,zirakashvili08}. It 
decreases the diffusion coefficient and increases the maximum energy of accelerated particles. 
The results of continuing theoretical study of this effect can be found in review papers 
\cite{bell2014,Caprioli2014}. 

In our calculations below, we use the Bohm value of the 
diffusion coefficient $D_B=pvc/3qB$, where $q$ is the electric charge of particles. 
The amplified upstream magnetic field is $B=V_f\sqrt{4\pi \rho}/M_A$.
Here $V_f$ is the speed of the forward shock.
The value of the parameter $M_A=10$ describing the magnetic field amplification is assumed.

Figures (1)-(2) illustrate the results of our numerical calculations.

We stop our calculations at $t=30$ yr when the stellar wind mass swept up by the forward shock is of the
order of
 20 $M_\odot$. This value of the total mass loss is expected if the initial mass of the Type IIn supernova progenitor
 is $60-70M_{\odot}$. The rest goes into ejecta ($\sim $ $10M_\odot $) and the black hole ($\sim $ $40M_\odot $).
 At later times the forward shock enters rarefied medium created by the fast tenuous wind of
 supernova progenitor at the main sequence stage. So the production of neutrinos becomes negligible at this time.

\begin{figure}[t]
\begin{center}
\includegraphics[width=8.0cm]{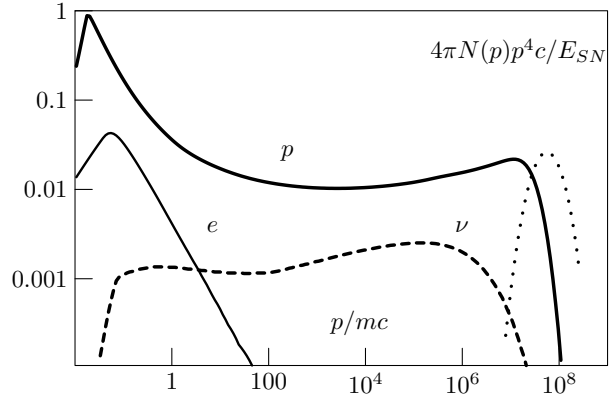}
\end{center}
\caption{ Spectra of particles produced in the supernova remnant
during $30$ yr after explosion. Spectrum of protons
(thick solid line ), spectrum of secondary electrons
 (multiplied on $10^3$, thin solid line), spectrum of neutrinos
(thick dashed line) are shown.}
\end{figure}

\begin{figure}[t]
\begin{center}
\includegraphics[width=8.0cm]{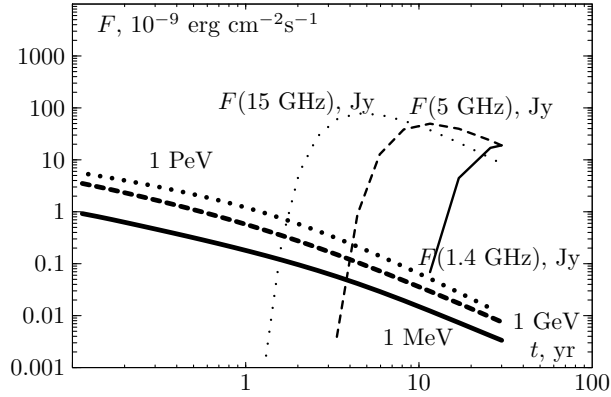}
\end{center}
\caption{Dependencies on time of fluxes from supernova remnant at distance $1$ Mpc. We show
 the  neutrino flux at 1 PeV produced via pion decay (thick dotted line),
gamma-ray flux at 1 GeV produced via pion decay
 (thick dashed line).
 The evolution of synchrotron gamma-ray flux at $1$ MeV (thick solid line) and
 radio-fluxes  at $1.4$ GHz, $5$ GHz and $15$ GHz  (thin solid, dashed and dotted lines respectively) are  also shown.}
\end{figure}

The spectra of particles produced during 30 years after explosion
are shown in Fig.1. They are calculated as the sum of the spectra
integrated throughout simulation domain and of the time-integrated
outward diffusive flux at the simulation boundary at $r=2R_f$
where $R_f$ is the forward shock radius. The maximum energy of
accelerated protons is about $30$ PeV, while higher energy
particles have already left the remnant. We also show the
time-integrated spectrum of neutrinos produced in $\it{pp}$
interactions. We checked that $p\gamma $ interactions result in
 the negligible contribution to neutrino fluxes for realistic supernova optical
  luminosity.

About $25\%$ of the kinetic energy of explosion goes to cosmic rays.

Evolution of non-thermal emission produced in the supernova remnant at distance
$1$ Mpc is shown in Fig.2. We take into account the synchrotron
self-absorption and free-free thermal absorption that are
 important for radio-supernovae \cite{chevalier98}. We use the temperature $T=10^4$K of circumstellar
 wind that is a characteristic value for dense stellar winds ionized by radiation coming from
the forward and reverse shocks \cite{lundqvist88}.

It is instructive to compare results of the present calculations with the approximate analytical
expression for the maximum energy of accelerated protons derived in our paper \cite{ptuskin10}.
The maximum energy of protons $E_{\max}$ accelerated by a supernova shock in the stellar wind was estimated as

\begin{equation}
E_{\max }=8\ \mathrm{PeV } \left( \frac {\dot{M}}{10^{-4}\ M_{\odot}\ \rm{yr}^{-1}}\right) ^{1/2}
\left( \frac {u_w}{100\ \rm{km\ s}^{-1}}\right) ^{-1/2}
\left( \frac {E_{SN}}{10^{52}\ \rm{erg}}\right)
\left( \frac {M_{ej}}{10M_{\odot}}\right) ^{-1}
\end{equation}

 A similar estimate for Type Ia supernova explosion in the uniform medium gives
the "knee" energy $\sim 3$ PeV for protons. The shocks propagating in stellar
 winds accelerate particles to higher energies. In particular Type IIn
supernova remnants with their dense winds with mass-loss rate $\dot{M}
\sim 10^{-2}\ M_{\odot}$ yr$^{-1}$ can accelerate protons up to $80$
PeV according
 to Eq. (2.2). This order of magnitude estimate is in agreement
with our numerical results illustrated in Fig. 1. Thus, explanation
of proton acceleration up to $\sim 10^{17}$ eV in Type IIn supernova
remnants is in line with the acceleration of cosmic rays up to the
knee in Type Ia supernova remnants. The "knee" energy of accelerated
particles is reached at the beginning of the Sedov stage when a significant
 fraction of explosion energy  transfers to the shock wave.
In denser wind, the amplified magnetic field is stronger that leads to larger $E_{\max }$.

\section{Calculation of background neutrino flux}

The neutrino spectrum produced by a single supernova remnant $Q(E_{\nu})$ (see Fig.1) can be used
for the determination of
extragalactic neutrino background. Distributed in the Universe Type IIn supernova remnants give the
following diffuse flux of neutrinos:

\begin{equation}
F(E_{\nu})=\frac {c}{4\pi H_0}\int ^{z_{max}}_{0}dz \frac {Q((1+z)E_{\nu})\nu_{sn}(1+z)^{m}}
{\sqrt{\Omega_{m}(1+z)^{3}+\Omega_{\Lambda}}}
=\frac {c}{4\pi H_0}\int^{(1+z_{max})E_{\nu}}_{E_{\nu}}dE'\frac{E'^{m}}
{E_{\nu}^{m+1}}\frac {\nu_{sn}Q(E')}{\sqrt{\Omega_{m}E'^{3}/E_{\nu}^{3}+\Omega_{\Lambda}}}.
\end{equation}

Here the adiabatic energy loss of neutrinos produced at the redshifts $0\leq z \leq z_{max}$ is taken
into account. The present neutrino production rate per unit energy and volume is $\nu_{sn}Q(E_{\nu})$,
where $\nu_{sn}$ is the rate of Type IIn supernovae at $z=0$ while the cosmological evolution of the
sources in the comoving volume is described as $(1+z)^{m}$ ($m=0$ implies no evolution).
The evolution
parameter $m=3.3$ for $z<1$ and no evolution at $z>1$, the maximum redshift $z_{\max }=5$ and the
rate $\nu _{sn}=10^{-6}$ Mpc$^{-3}$ yr$^{-1}$ at $z=0$ are assumed in our calculations
(see e.g. \cite{dahlen12}).
This rate of Type IIn supernovae is $100$ times lower than the rate of all core collapse
supernovae. $H_{0}=70$ km s$^{-1}$ Mpc$^{-1}$ is the Hubble parameter at the present epoch,
the matter density in the flat Universe is $\Omega_{m}=0.28$, and the $\Lambda$ -term is $\Omega_{\Lambda}=0.72$.

The calculated background neutrino spectrum is shown in Fig.3. The figure demonstrates very good fit
of our calculations to the IceCube data.

\begin{figure}[t]
\begin{center}
\includegraphics[width=8.0cm]{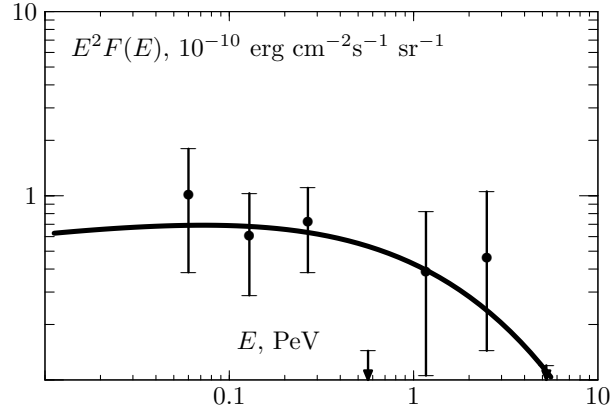}
\end{center}
\caption{ Calculated spectra of neutrino produced by IIn SNRs in the expanding Universe (solid line).
IceCube data \cite{aartsen14} are also shown. }
\end{figure}

Using the same approach we can calculate the flux of extragalactic protons. Our results are compared
 with cosmic ray data in Fig.4. So the proton flux produced by  extragalactic IIn supernova is below
 the measured cosmic ray fluxes.

The possible contribution of Galactic Type IIn SN events to the observed
cosmic ray intensity might be comparable with the Type IIb supernovae as
it was calculated in our paper [11]. Both types of supernovae probably
produce approximately the same amount of cosmic rays with almost the
same maximum energy of particles but SNIIn generate much more secondary
particles and neutrinos. The low expected rate of these Galactic supernovae,
about one per 5 thousand years, makes estimates of their contribution to
the observed intensity of  very high energy Galactic cosmic rays rather uncertain.

Simple order of magnitude estimates can be done to clarify how the obtained neutrino
flux depends on supernova parameters. The main production of high energy particles
and neutrinos is finished at the beginning of the Sedov stage when the shock radius $R_{S}$
can be determined from the condition
$M_{ej}=4\pi \int^{R_{S}} drr^{2}\rho=\dot{M}R_{S}/u_{w}$. The spectrum $E^{-2}$ is
assumed for the accelerated particles.
The neutrino energy flux expected from a supernova at distance $D$ can be estimated as
(see also \cite{kats11,murase14})

\[
f(E_{\nu})E_{\nu}^2=\frac {3\xi _{CR}K_{\nu }}{16\pi ^2\ln (E_{\max
}/mc^2)}\frac {cV_f^2\sigma _{pp}\dot{M}^2}{mu^2_wR_fD^2}=
\]
\begin{equation}
10^{-8} \rm{erg\ cm}^{-2}\rm{s}^{-1} t_{yr}^{-1}D_{\rm{Mpc}}^{-2}\xi_{CR}
\left( \frac {\dot{M}}{10^{-2}\ M_{\odot}\ \rm{yr}^{-1}}\right) ^{2}
\left( \frac {u_w}{100\ \rm{km\ s}^{-1}}\right) ^{-2}
\left( \frac {E_{SN}}{10^{52}\ \rm{erg}}\right) ^{1/2}
\left( \frac {M_{ej}}{10\ M_{\odot}}\right) ^{-1/2}
\end{equation}
Here $\xi _{CR}$ is the ratio of cosmic ray pressure to the ram
pressure of the shock $\rho V_f^2$, $K_\nu \approx 0.25$ is the
fraction of energy that goes into  neutrinos in {\it pp}
interactions, $\sigma _{pp}$ is the cross-section of {\it pp}
interactions
and $E_{\max }$ is the maximum energy of accelerated protons given by Eq. (2.2). The expression is valid when the
 shock radius $R_f>R_{\min }=c\sigma _{pp}\dot{M}/4\pi u_wmV_f$ and the energy losses via {\it pp}
interactions are smaller than the adiabatic losses.

Expression (3.2) can be used for the determination of neutrino background produced by all Type IIn supernovae:

\[
F(E_{\nu})E_{\nu}^2=\frac {3\xi _{CR}K_{\nu }}{16\pi ^2\ln (E_{\max }/mc^2)}
\frac {\nu _{sn}c^2V_f\sigma _{pp}\dot{M}^2}{H_0mu^2_w}\ln{\frac {R_S}{R_{\min }}}
=10^{-11} \rm{erg\ cm}^{-2}\rm{s}^{-1} \rm{sr}^{-1}\times
\]
\begin{equation}
\xi_{CR}
\left( \frac {\dot{M}}{10^{-2}\ M_{\odot}\ \rm{yr}^{-1}}\right) ^{2}
\left( \frac {M_{ej}}{10\ M_{\odot}}\right) ^{-1/2}
\left( \frac {u_w}{100\ \rm{km\ s}^{-1}}\right) ^{-2}
\left( \frac {E_{SN}}{10^{52}\ \rm{erg}}\right) ^{1/2}
\left( \frac {\nu _{sn}}{10^{-6} \rm{Mpc}^{-3}\rm{yr}^{-1}}\right) ,
\end{equation}
The background  flux is higher if  the cosmological evolution
of supernova rate is taken into
account (see Eq. (3.1)).
These estimates are valid at neutrino energies $E_{\nu}<0.05E_{\max }$. They confirm the efficiency
of high energy neutrino production in supernova explosion in very dense wind of a progenitor star,
as it takes place in  Type IIn supernovae.

\section{Discussion}

Our calculations show that the IceCube data can be explained by the
 neutrinos from Type IIn supernovae. If so, the arrival direction of every IceCube
neutrino should coincide with the direction
 to some Type IIn supernova. However, we observe Type IIn supernovae
with the redshifts $z<0.1$ while the background is determined by the
 supernovae with $z\sim 1$. Therefore we expect that only several percent
of observed neutrinos i.e. about one IceCube neutrino event are
associated with the known Type IIn supernovae. Unfortunately,
the arrival directions of majority of IceCube neutrinos are known with low
angular resolution (about 15 degrees).
We found that several Type IIn supernovae were indeed observed
in the vicinity of some IceCube neutrinos.
However this result has low statistical significance.

Nevertheless we note that the arrival direction of one IceCube PeV neutrino candidate (event 20) is
 within 5 angular degrees from the SN 1978K. This  nearest IIn supernova at distance 4 Mpc is for
 decades observed in radio, X-rays and optics (e.g. \cite{smith07}). The mass-loss rate of SN 1978K progenitor is
 estimated  as $\dot{M}=2\cdot 10^{-3}\ M_{\odot}$ yr$^{-1}$ \cite{chugai95} that is 5 times lower than we use in our
 calculations. Thus the fluxes expected from this supernova are not so high as ones shown in Fig.2.
 In spite of this it is possible that the PeV neutrino candidate was emitted by SN 1978K. 

The arrival direction of another IceCube PeV neutrino candidate (event 35) is
 within 10 angular degrees from the SN 1996cr. This  nearest IIn supernova at distance 4 Mpc was also 
 detected in radio, X-rays and optics (e.g. \cite{bauer08}). It is known that the forward shock of 
 SN1996cr entered the dense shell of circumstellar matter. The radio flux of this supernova is only 
 several times lower than the flux shown in Fig.2. That is why this IceCube PeV neutrino candidate 
 could be emitted by SN 1996cr. 

 On the other hand,  $8$ of $37$ IceCube neutrinos left tracks. The arrival directions of these neutrinos
are determined with good
 angular resolution (about 1 degree). We found that none of these neutrinos is associated
with known Type IIn supernovae. The expected number of coincidences is of the order of $0.3$.
Thus, only one track neutrino can be observed from the vicinity of Type IIn supernova during
one decade of IceCube operation.

\section{Conclusions}

Our main conclusions are the following:

1) The diffusive shock acceleration of particles in Type IIn supernova remnants and the production
of neutrinos via $pp$ interactions in the dense presupernova winds can explain the
diffuse flux of high energy astrophysical neutrinos observed in the IceCube experiment.

2) The calculated maximum energy of protons accelerated in
remnants of IIn supernova is close to $10^{17}$ eV. This value is
higher than the maximum energy achieved in the main part of SNRs
and is explained by the high density of the circumstellar matter.

3) The efficient  acceleration of particles and production of secondary electrons and positrons result in the fluxes
of radiowaves and gamma-rays that can be observed from the
nearest Type IIn supernova remnants.

4) Future IceCube operation and the search of correlations between neutrino arrival directions
 and the directions to Type IIn supernovae may check whether these objects are the
sources of high energy neutrinos but only one track event during $10$ years is expected
from these supernovae with the redshifts less than $0.1$. The bright phase of a
Type IIn supernova remnant as a source of PeV neutrinos lasts for about several years after the supernova explosion.

The work was supported by the Russian Foundation for Basic Research
grant 13-02-00056 and by the Russian Federation Ministry of Science
and Education contract 14.518.11.7046.

\begin{figure}
\begin{center}
\includegraphics[width=8.0cm]{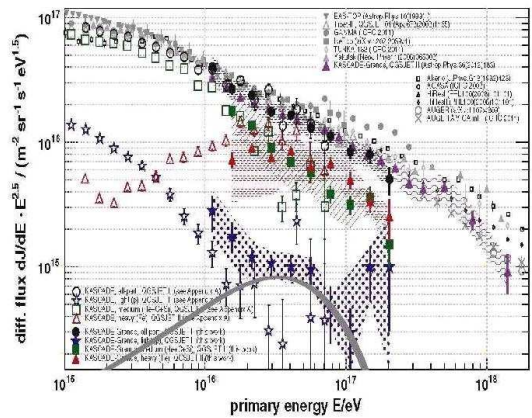}
\end{center}
\caption{ Comparison of calculated spectrum of cosmic ray protons produced by extragalactic IIn SNRs
 in the expanding Universe (gray solid line) and available cosmic ray data \cite{apel13}.}
\end{figure}


\begin{thebibliography}{99}

\bibitem{aartsen13} Aartsen, M.G., Ackermann, M., Adams, J. et al. 2013,
Science 342, 1242856

\bibitem{aartsen14} Aartsen, M.G., Ackermann, M., Adams, J. et al. 2014,
Phys. Rev. Lett., 113, 1101

\bibitem{aartsen14b} Aartsen, M.G., Ackermann, M., Adams, J. et al. 2014,
ApJ, 796, 109

\bibitem{Anchor14} Anchordoqui, L.A., Barger, V., Cholis, E. et al. 2014, JHEAp, 1, 1

\bibitem{GaisserHalz14} Gaisser, T. \& Halzen F. 2014, Annu. Rev. Nucl. Part. Sci. 64, 101

\bibitem{murase14a} Murase, K 2014, arXiv:1410.3680v2

\bibitem{krymsky77} Krymsky, G.F. 1977, Soviet Physics-Doklady, 22, 327

\bibitem{bell78} Bell, A.R., 1978, MNRAS, 182, 147

\bibitem{axford77} Axford, W.I., Leer, E. \& Skadron, G., 1977, Proc. 15th
Int. Cosmic Ray Conf., Plovdiv, 90, 937

\bibitem{blandford78} Blandford, R.D., \& Ostriker, J.P. 1978, ApJ, 221, L29

\bibitem{ptuskin10} Ptuskin, V.S., Zirakashvili, V.N. \& Seo, E.S. 2010, ApJ, 718, 31

\bibitem{moriya14} Moriya, T.J., Maeda, K., Taddia, F., Sollerman, J., Blinnikov, S.I., \&
Sorokina, E.I., 2014, MNRAS 439, 2917

\bibitem{kats11} Kats, B., Sapir, N., \& Waxman, E., 2011, arXiv.1106.1898

\bibitem{murase14} Murase, K., Thompson, T.A., \& Ofek, E.O., 2014, MNRAS 440, 2528

\bibitem{zirakashvili12} Zirakashvili, V.N. \& Ptuskin V.S. 2012, APh 39, 12

\bibitem{berezhko94} Berezhko, E.G., Elshin, V.K. \& Ksenofontov, L.T., 1994, APh 2, 215

\bibitem{kang06} Kang, H. \& Jones, T.W. 2006, APh 25, 246

\bibitem{berezhko07} Berezhko, E.G., \& V\"olk, H.J., 2007, ApJ, 661, L175

\bibitem{bell04} Bell, A.R., 2004, MNRAS, 353, 550

\bibitem{zirakashvili08} Zirakashvili, V.N. \& Ptuskin, V.S., 2008, ApJ 678, 939

\bibitem{bell2014} Bell, A.R., 2014, APh 43, 56

\bibitem{Caprioli2014} Caprioli, D., 2014, arXiv:1410.1644v1



\bibitem{chevalier98} Chevalier, R., 1998, ApJ 499, 810

\bibitem{lundqvist88} Lundqvist, P., \& Fransson, C., 1988, A\&A 192, 221

\bibitem{dahlen12} Dahlen, T., Strolger, L.-G., Riess, A.G. et al., 2012, ApJ, 757, 70

\bibitem{apel13} Apel, W.D., Arteaga-Vel\'azquez, J.C., Bekk, K. et al. 2013, APh 47, 54

\bibitem{smith07} Smith, I.A., Ryder, S.D., B\"ottcher, M., et al., 2007, ApJ, 669, 1130

\bibitem{chugai95} Chugai, N.N., Danziger, I.J., \& Della Valle, M., 1995, MNRAS, 276, 530

\bibitem{bauer08} Bauer, F.E., Dwarkadas, V.V., Brandt, W.N. et al., 2008, ApJ, 688, 1210



\end{thebibliography}
\end{document}